\documentstyle[lprocl,11pt,epsf]{article}

\def\be{\begin{equation}}
\def\ee{\end{equation}}
\def\bea{\begin{eqnarray}}
\def\eea{\end{eqnarray}}

\newcommand{\lsim}{\raisebox{-0.6ex}{
                        $\stackrel{\textstyle <}{\textstyle\sim}$}}
\def\pslash{/\!\!\!\!p}
\def\Dslash{/\!\!\!\!D}
\def\qslash{/\!\!\!\!q}
\def\vslash{/\!\!\!\!v}

\begin{document}

\vbox{\hbox{\rightline{CALT-68-2130}}\hbox{\rightline{hep-ph/9707522}}\hbox{}}

\title{RECENT PROGRESS IN EFFECTIVE FIELD THEORY
\footnote{Talk given at the VII Blois Workshop on Hadronic Physics, 
Seoul, Korea.} 
}

\author{MARK B. WISE}

\address{California Institute of Technology, Pasadena, CA 91125, USA}

\maketitle

\abstracts{Applications of effective field theory to nucleon-nucleon
scattering, quarkonia decay and production and B meson decay are discussed. 
Some unresolved issues are considered.}
 
\section{Introduction}

The use of effective field theories is a standard tool for dealing with strong
interaction phenomena in the nonperturbative regime. The basic idea is to
construct the most general Lagrangian consistent with the symmetries of Quantum
Chromodynamics (QCD) out of fields that destroy the relevant degrees of freedom
in the problem. For example, in the chiral Lagrangian for pion self
interactions, those degrees of freedom are the pion fields, while in the heavy
quark effective theory (HQET) and nonrelativistic quantum chromodynamics
(NRQCD) they are quark and gluon fields. If this were all that was done one
would have no more predictive power than that based on the symmetries and
unitarity of the S-matrix. To endow the method with more predictability, there
must be a power counting scheme where the effective Lagrangian is expanded in a
small parameter to reduce the number of operators that occur in it. In the case
of chiral perturbation theory for pion interactions the small parameter is the
pion momentum divided by a typical hadronic scale.  For HQET the expansion
parameter is a typical hadronic scale divided by the heavy quark mass, and for
NRQCD the expansion is in the typical velocity of the heavy quarks divided by
the speed of light. Often the leading term in the expansion has new
(approximate) symmetries that were not manifest in the QCD Lagrangian. In
chiral perturbation theory (CPT) the approximate symmetry is the $SU(3)_L
\times SU(3)_R$ chiral symmetry that occurs when the light quark masses are
neglected. For HQET they are the heavy quark spin-flavor symmetries that occur
when the heavy quark masses are taken to infinity with their four velocities
fixed. In NRQCD there are also heavy quark spin symmetries.

At a practical level, the power counting takes somewhat different forms
depending on what type of regulator is used for the effective field theory.
Consider the case of the effective field theory for Cabibbo allowed nonleptonic
charm decays that results from integrating out the W-boson.  At tree level the
effects of the Feynman diagram in Fig.~1 are reproduced by the effective
Lagrangian
\begin{equation}
{\cal L}=C_1 O_1+C_2 O_2 +...,
\label{eq:eff}
\end{equation}
where
\begin{equation}
C_1=g_2^2/8M_W^2, \qquad C_2=g_2^2/8M_W^4.
\label{eq:2}
\end{equation}
The ellipses in Eq.~\ref{eq:eff} refer to operators with more derivatives.  
In Eq.~\ref{eq:2} $M_W$ is the W-boson mass and $g_2$ is the weak $SU(2)$ 
coupling.  The operators  $O_1$ and $O_2$ are
\begin{eqnarray}
O_1 &=& [\bar s \gamma_{\mu}(1-\gamma_5)c]\,
  [\bar u \gamma^{\mu}(1-\gamma_5)d],  \nonumber\\*
O_2 &=& [\bar s \gamma_{\mu}(1-\gamma_5)c]\,
  \Box [\bar u \gamma^{\mu}(1-\gamma_5)d]. 
\end{eqnarray} 
At tree level the effects of $O_2$ on the $c \rightarrow \bar d us$ decay rate
are suppressed by $p^2/M_W^2$, compared with those of $O_1$, where p is a
typical momentum in the decay.

\begin{figure}[t]
\centerline{\epsfysize=2truecm \epsfbox{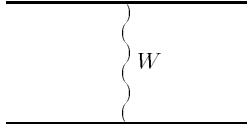} }
\caption{Virtual $W$-boson exchange diagram.}
\end{figure}

When perturbative order $\alpha_s$ corrections are considered it is necessary
to regulate the theory because of ultraviolet divergences coming from the large
momentum part of the loop integration in Feynman diagrams.  Imagine using a
momentum cutoff, $\Lambda$, for the regulator. The size of the contribution of
$O_2$ depends on the value of the cutoff. Its contribution, for $\Lambda^2 \sim
M_W^2$, is suppressed by $\alpha_s$ compared with that of $O_1$. It no longer
appears to be suppressed by $p^2$. This large part of its contribution can be
absorbed into a redefinition of the coefficient of $C_1$. If a momentum cutoff
is used as the regulator, the value of $C_1$ extracted from experiment in the
theory with $O_2$ included may differ from its value in the theory without
$O_2$ by an amount of order  $\alpha_s$. Nonetheless the net new effect on the
physics due to $O_2$ is suppressed by $p^2/M_W^2$. Even with a dimensionful
cutoff there is a subtraction procedure that has all the effects of $O_2$
suppressed by $p^2/M_W^2$. This occurs if renormalized operators are defined so
that, for the renormalized version of $O_2$ (i.e. $O_2^R$), the part of its
matrix elements that grow as $\Lambda^2$ and are proportional to matrix
elements of $O_1$ are subtracted away. For the Lagrangian, ${\cal L}=C_1^R
O_1^R+C_2^R O_2^R +... $, the extracted value of $C_1^R$ at order $\alpha_s$
will not depend on whether $O_2^R$ is included or not since all its effects are
of suppressed by $p^2/M_W^2$. The value of the cutoff can be arbitrarily large.
However, if such renormalized operators are not introduced for $\Lambda^2 \gg
M_W^2$ the ``physical''  $c \rightarrow \bar d us$ amplitude will arise from a
delicate cancellation between the contribution of $C_1$ and the one loop
contribution of $O_2$ proportional to $\alpha_s \Lambda^2/M_W^2$ .

Using dimensional regularization (with mass independent $\overline {\rm MS}$
subtraction) is a little like taking the momentum cutoff to infinity. But with
this regulator power divergences are automatically subtracted and the effects
of $O_2$ are manifestly suppressed by $p^2/M_W^2$.   Using almost any type of
ultraviolet regulator is fine. However, one might encounter technical issues
with one type of regulator that are not there with another, for example adding
counter terms to restore gauge invariance in the case of a momentum cutoff. For
the remainder of this lecture, I assume that dimensional regularization with
minimal subtraction is used.

My lecture will be divided into five Sections. In Section 2 I discuss an issue
in the application of chiral perturbation to nucleon-nucleon scattering that
needs to be resolved before this can be claimed to be a useful systematic
method.  Section 3 discusses some recent applications of NRQCD that show that
terms that are naively suppressed by powers of $v/c$ can become important in
certain kinematic regions. Section 4 deals with recent developments in NRQCD
formalism and finally Section 5 briefly describes a recent application of HQET
to semileptonic B decay to excited charmed  mesons.

\section{Chiral Perturbation Theory for Nucleon-Nucleon Scattering}

Weinberg first suggested using CPT for nuclear physics.~\cite{weinberg} The
basic idea was that a power counting can be established for the nucleon-nucleon
potential which is then used to calculate properties of systems of nuclei, e.g.
$NN$ phase shifts.  (For a review of some applications see Ref.~[2].)  In the
$^1S_0$ channel the leading nucleon-nucleon potential is supposed to be

\begin{equation}
V_0({\bf p},{\bf p'})=\tilde C-4\pi \alpha_{\pi}/({\bf q}^2+m_{\pi}^2),
\label{eq:potential}
\end{equation}
where ${\bf p}$ and ${\bf p'}$  are the relative three momenta of
the initial and final $NN$ pair, ${\bf q}={\bf p}-{\bf p'}$, and

\begin{equation}
\alpha_{\pi}= g_A^2m_{\pi}^2/8{\pi}^2f_{\pi}^2.
\label{eq:nuc}
\end{equation}
In Eq.~\ref{eq:nuc} $g_A\simeq 1.25$ is the axial current coupling and
$f_{\pi}\simeq 132$MeV is the pion decay constant. The second term in
Eq.~\ref{eq:potential} comes from one pion exchange.  You might not recognize
its form because pions are derivatively coupled. In Eq.~\ref{eq:potential} the
numerator was arrived at by writing ${\bf q}^2$ as $({\bf
q}^2+m_{\pi}^2)-m_{\pi}^2$. The first of these two terms gives a constant in
the potential that was absorbed into $\tilde{C}$ by the definition, $\tilde C=
C+g_A^2/2f_{\pi}^2$. The constant $C$ is the coefficient of a four nucleon
operator of the form $N^{\dagger}NN^{\dagger}N $. Note that four nucleon
operators containing derivatives or insertions of the quark mass matrix are
supposed to give contributions to the potential that are suppressed.

\begin{figure}[h]
\centerline{\epsfysize=2truecm \epsfbox{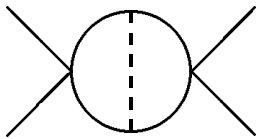} }
\caption{Two loop diagram contributing to $^1S_0$ nucleon-nucleon scattering.
The dashed line represents virtual pion exchange.}
\end{figure}

Imagine using the potential in Eq.~\ref{eq:potential} to calculate the $^1S_0$
$NN \rightarrow NN$ phase shift. This is done by solving the Schrodinger
equation or equivalently summing the ladder diagrams. The potential is singular
(in coordinate space the constant term in $V_0({\bf p},{\bf p'})$ corresponds
to a delta function potential) and ultraviolet divergences are encountered. For
example Fig.~2 gives rise to a divergent amplitude proportional to
\begin{equation}
{1 \over d-4}\tilde C^2\alpha_{\pi}M_N^2/4\pi,
\label{eq:diverge}
\end{equation}
where $d$ is the space-time dimension. (If a cutoff were used instead of
dimensional regularization then the factor of $1/(d-4)$ would be replaced by
$ln(m_{\pi}^2/\Lambda^2)$.) The divergence in Eq.~\ref{eq:diverge} is removed
by adding a counter term.  The subtraction point dependence associated with the
finite part of Fig.~2 is canceled by the subtraction point dependence in the
coefficient of the counter term operator. In this case the counter term is of
the form~\cite{kaplan} 
\begin{equation}
{\cal L}_{ct} \sim C_{ct} Tr(\Sigma {m_q}+\Sigma^{\dagger} 
  {m_q})N^{\dagger}NN^{\dagger}N .
\label{eq:counter term}
\end{equation}

Eq.~\ref{eq:counter term} is necessary as a counter term for the leading order
calculation. The only possible power counting prescription that could make its
contribution to the potential subdominant to that in Eq.~\ref{eq:potential}
would be one based on considering logarithms of $(m_q/\Lambda_{QCD})$ as large.
This situation is similar to what happens in pion self interactions. There as
one includes higher and higher loops, terms with more insertions of the quark
mass matrix and/or higher derivatives must also be included in the chiral
Lagrangian . (Actually in the case of $NN$ $^1S_0$ scattering, no divergences
corresponding to operators with derivatives occur. However the work of Ref.~[4]
makes is seem unlikely that a reasonable power counting can be developed for
which such operators are subdominant.) 

The problem of elastic nucleon-nucleon scattering itself is not that
interesting. However, if a systematic approach using CPT is possible for this
problem, then one can contemplate using it also for inelastic pion production
and for the study of nuclear matter. It seems worth exploring further whether a
systematic approach to nucleon-nucleon scattering based on CPT is possible.

\section{Kinematically Enhanced Nonperturbative Corrections in NRQCD}

NRQCD is an effective field theory used to predict properties of quarkonium in
an expansion in $v$, where $v$ is the magnitude of the relative velocity of the
heavy $\bar Q Q$ quark pair. (In this section we adopt the usual particle
physics convention that the speed of light is $c = 1$.) It has made surprising
predictions for quarkonia production and decay because effects suppressed by
powers of $v$ can be enhanced by factors of $1/\alpha_s(m_Q)$ compared with the
leading term in the $v$ expansion. As an example of this phenomena consider the
inclusive decay $\Upsilon \rightarrow \gamma+X$, where $X$ denotes light
hadrons. The differential decay rate can be written as the imaginary part of a
time ordered product of electromagnetic currents (there are some complications
due to overlapping unwanted cuts, see Ref.~[5] for a discussion of this.),
\begin{equation}
{d\Gamma \over dE_{\gamma}}={2e_b^2 \over \pi} E_\gamma ImT',
\end{equation}
where $e_b=-1/3$ is the b-quark charge and $E_\gamma$ is the photon energy. The
operator product expansion (OPE) can be used to calculate the imaginary part of
the time ordered product $T'$. For a given photon energy the final hadronic
invariant mass squared is $m_X^2=m_{\Upsilon}^2(1-2E_{\gamma}/m_{\Upsilon})$ so
near the endpoint $E_{\gamma}=m_{\Upsilon}/2$ only low mass hadronic final
states can occur. In this region predictions based on the OPE must be smeared
over a region of photon energies, $\Delta E_{\gamma}$, before they can be
compared with experiment.  If the smearing region is chosen too small, higher
dimension operators in the OPE become successively more important and the OPE
breaks down. In this section we examine the type of operators that control this
endpoint region and also show that analogous operators play an important role
in predictions for quarkonium production based on NRQCD. 

\begin{figure}[t]
\centerline{\epsfysize=3truecm \epsfbox{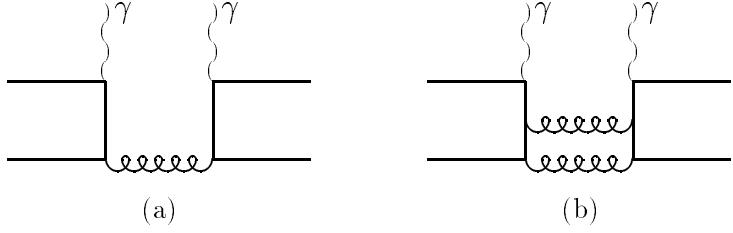} }
\caption{Feynman diagrams contributing to ${\rm Im}T'$.}
\end{figure}

At tree level the OPE is performed by calculating the tree level Feynman
diagrams in Figs.~3. Fig.~3a gives a contribution of order $\alpha_s(m_b)$,
while Fig.~3b gives a contribution of order $\alpha_s(m_b)^2$. The  order
$\alpha_s(m_b)$ contribution to $T'$ from Fig.~3a, denoted by $T'_8$, is called
the octet contribution. It is usually neglected because it is higher order in
$v$ than the color singlet contribution from Fig.~3b (which we denote by
$T'_1$). However $T'_8$ is enhanced by $1/\alpha_s(m_b)$ compared with $T'_1$.
The Feynman diagram in Fig.~3a gives 

\[
T'_8=4g^2 \langle \Upsilon|{{\left[ \bar b\gamma^{\alpha}(\pslash +i\Dslash -\qslash +m_b)T^A\gamma_{\nu} b\right]}\over {(p+iD-q)^2-m_b^2}}  
 {g^{\mu\nu} \over {(2p-q+iD)^2+i\epsilon}} \]
\begin{equation}
\times {{\left[ \bar b\gamma_{\mu}(\pslash +i\Dslash -\qslash +m_b)T^A\gamma_{\alpha} b\right]}\over {(p+iD-q)^2-m_b^2}}|\Upsilon \rangle .
\label{eq:T_8}
 \end{equation} 
In Eq.~\ref{eq:T_8} $g$ is the strong coupling, $D$ denotes a covariant
derivative, $p=(m_b,{\bf 0})$ and the Upsilon states are at rest. The imaginary
part comes from the gluon propagator which produces the factor $\delta
((2p-q+iD)^2)$. Expanding this in $D$ gives a sequence of operators that are
more and more singular in the endpoint region.  

Making the transition to NRQCD there are two contributions depending on the 
spin structure of the $\bar b b$ pair,
\begin{equation}
T'_8=T'_8(^3P_J)+T'_8(^1S_0),
\end{equation}
and the expansion in $D$ of Eq.~\ref{eq:T_8} gives

\begin{eqnarray}
ImT'_8(^1S_0)={{7\pi g^2} \over {18m_b^5}}\sum_{m=0}^{\infty}{{(-1)^m} \over {m!}}\langle \Upsilon|\left[\psi^{\dagger}
T^A \chi \right](in \cdot D)^m \left[\chi^{\dagger} T^A \psi \right]|\Upsilon \rangle \delta^{(m)} (2E_{\gamma}-2m_b).
\label{eq:sum}
\end{eqnarray}
In Eq.~\ref{eq:sum} $n=(2p-q)/m_b$, which can be taken to be equal to
$(1,0,0,1)$, is a light-like four-vector. The superscript $m$ on the delta
function denotes its m'th derivative with respect to $2E_{\gamma}$ and the
$\psi$ and $\chi^{\dagger}$ are the two component Pauli spinor fields that
destroy the bottom quark and antiquark in NRQCD. A similar expression holds for
$ImT'_8(^3P_J)$.

The $m$'th term in the sum of Eq.~\ref{eq:sum} scales as $v^{7+2m}$ according
to the NRQCD counting rules. Here we are assuming that $\Lambda_{QCD}/m_b$ is
of order $v^2$. The contribution to the total $\Upsilon \rightarrow \gamma+X$
rate is dominated by the $m=0$ term which gives a contribution of order
$(\alpha_s(m_b) / \pi )v^7$. Note that the color singlet contribution to the
decay rate is of order $(\alpha_s(m_b)/ \pi )^2v^3$. If we take $\alpha_s(m_b)
/ \pi $ to be of order $v^2$, then the color singlet term dominates the rate.
But the octet contribution is very important near the endpoint. If we smear the
endpoint region of the differential decay rate over photon energies $\Delta
E_{\gamma}<<m_bv^2$, then successive terms in the sum over $m$ in
Eq.~\ref{eq:sum} become more important. On the other hand if we smear the
differential decay rate over a region of photon energies $\Delta
E_{\gamma}>>m_bv^2$, then successive terms in the sum over $m$ in
Eq.~\ref{eq:sum} become less important and the color singlet contribution
dominates. For $\Delta E_{\gamma} \sim m_bv^2$ all terms in the sum are of
comparable importance to the color singlet contribution. Similar remarks hold
for the contribution from $ImT^{\prime}_8(^3P_J)$. Note that the region of
photon energies, $\Delta E_{\gamma} \sim m_bv^2$ corresponds to a range of
final hadronic masses $\Delta m_X \sim m_b v$ which is much greater than the
QCD scale. Similar remarks hold for $\psi$ decay.

An analogous calculation in the color singlet case gives a sum of the form 
\begin{equation}
ImT'_1=g^4G(E_{\gamma})\sum_{m=0}^{\infty}{{(-1)^m} \over {m!}}\langle \Upsilon|\left[\psi^{\dagger}
\sigma_k \chi \right](in \cdot \partial)^m \left[\chi^{\dagger} \sigma_k \psi \right]|\Upsilon \rangle \theta^{(m)} (2m_b-2E_{\gamma}).
\label{eq:suma}
\end{equation}
In Eq.~\ref{eq:suma} $G(E_{\gamma})$ is a smooth function of the photon energy
and the superscript $m$ on the theta function denotes its m'th derivative with
respect to $2E_{\gamma}$. Inserting a complete set of states between the
bilinears of NRQCD fields, one finds that the vacuum state dominates and $in
\cdot \partial $ produces a factor of the binding energy. Consequently
Eq.~\ref{eq:suma} becomes

\begin{equation}
ImT'_1=g^4G(E_{\gamma})\langle \Upsilon|\left[\psi^{\dagger}
\sigma_k \chi \right]\left[\chi^{\dagger} \sigma_k \psi \right]|\Upsilon \rangle \theta (m_{\Upsilon}-2E_{\gamma}).
\end{equation}
In the color singlet case the sum of singular terms has converted the parton 
kinematics to hadron kinematics. 

The endpoint region of the photon energy spectrum in $\Upsilon \rightarrow
\gamma +X$ decay is determined by a sum of leading twist
operators.\cite{rothstein} The situation is very analogous to the endpoint
region of electron energies in semileptonic $B$ meson decay.\cite{neubert} A
similar phenomenon also occurs in quarkonium production.\cite{beneke} For
example, at large transverse momentum $p_\perp$ quarkonium production at the
Tevatron is controlled by gluon fragmentation. The differential cross section
for $\psi$ production is

\begin{equation}
{{d \sigma} \over dp_{\perp}}= \int_{\hat z_{min}} {{d \hat z} \over {\hat z}} K(\hat z,p_{\perp})D_{g \rightarrow \psi}(\hat z,p_{\perp}).
\label{eq:frag}
\end{equation}
In Eq.~\ref{eq:frag}, $\hat z=p_+/k_+$ , where $k$ is the four-momentum of the
fragmenting gluon and $p$ is the four-momentum of the $c\bar c$ quark pair in
the $\psi$ (i.e., in the rest-frame of the $\psi$, $p=(2m_c,{\bf 0})$). The hat
is placed on $z$ because it is defined in terms of parton kinematics instead of
hadron kinematics. The function $K$ depends on the parton cross section and
parton distributions. In the region $p_\perp \sim 15$GeV, $K \sim \hat z^5$ so
the integral in Eq.~\ref{eq:frag} weights the fragmentation function $D_{g
\rightarrow \psi}(\hat z,p_{\perp})$ towards $\hat z =1$.

Including the most singular terms as $\hat z \rightarrow 1$ the gluon 
fragmentation function is

\begin{eqnarray}
D_{g \rightarrow \psi}(\hat z,2m_c) &=& 
  {{\pi \alpha_s(2m_c)} \over {24m_c^3}}\sum_m {1 \over {m!}} 
  \delta^{(m)}(1-\hat z)  \\*
&\times& \sum_X \langle 0|\left[ \chi^{\dagger} \sigma_i T^A \psi \right]
  |\psi +X \rangle \langle \psi +X|(in \cdot D/(2m_c))^m \left[ \psi^{\dagger} 
  \sigma_i T^A \chi \right]|0\rangle. \nonumber
\label{eq:15}
\end{eqnarray} 

In this case the light-like vector is $n=(1,0,0,-1)$ and the superscript $m$ on
the delta function denotes its $m$'th derivative. Terms in  the sum are of
order $v^{7+2m}$ and so if one focuses on a region near the endpoint $\hat
z=1$, of size $\Delta z \sim v^2$, then all terms in the sum are of comparable
importance. One crude way to gauge the importance of the higher order terms is
to imagine that their effect is to shift the delta function at $\hat z = 1$ to
a delta function at $\hat z =1-{\cal O}(v^2)$. Then with $K\sim \hat z^5$ we
see that there is a correction to the leading $m=0$ term in the sum of order
$4v^2 \sim 1$. Of course this is just dimensional analysis. Without a more
precise estimate of the higher order terms in the sum, we cannot be confident
that they are numerically important for $\psi$ production at the Tevatron at
large $p_{\perp}$. Similar sums occur in all quarkonia production processes and
are important near the boundary of phase space where there is sensitivity to
the difference between parton and hadron kinematics.  

\section{Effective Lagrangian for NRQCD}

NRQCD organizes contributions to the physical properties of quarkonia as an
expansion in powers of $v/c$. Clearly the limit of QCD that is appropriate in
this case is the large $c$ limit.  The original formulation of Bodwin, Braaten,
and Lepage~\cite{bodwin} was similar in some ways to the effective field theory
for Cabibbo allowed charm decay that resulted from integrating out the W-boson
using a momentum cutoff, $\Lambda \sim M_{W}c$, as the regulator for the
ultraviolet divergences.  In this formulation of NRQCD with a dimensionful
ultraviolet regulator, a given operator contributes at many different orders in
$v/c$, unless a subtraction procedure is adopted to remove the subdominant
pieces.  There is nothing wrong with such a formulation.  However, recently
NRQCD has been reformulated using dimensional regularization so that a given
operator automatically only contributes at a fixed order in $v/c$, much like
the usual formulation of HQET where a given operator contributes at a fixed
order in $\Lambda_{QCD}/m_Q$~\cite{grinstein,luke2}. In this section I review
this recent work. Unlike the other sections of this talk, factors of $c$ will
be explicit here.

Consider the QCD Lagrangian for gluons interacting with a heavy quark $Q$
\begin{equation}
{\cal L}_{QCD} = - {1\over 4} G_{\mu\nu}^B G^{B\mu\nu} + c Q (i\Dslash - m_Q c) Q. 
\label{eq:16}
\end{equation}
In Eq.~\ref{eq:16} the $0$ component of a partial derivative is
\begin{equation}
\partial_0 = {1\over c} \left({\partial\over\partial t}\right),
\end{equation}
and $D$ is the covariant derivative
\begin{equation}
D_\mu = \partial_\mu + {ig\over c} A_\mu^B T^B.
\label{eq:17}
\end{equation}
The gluon field strength tensor $G_{\mu\nu}^B$ is defined in the usual way except that $g \rightarrow g/c$.  

There are many choices for how factors of the speed of light $c$, are put into
Eq.~\ref{eq:16}.  This occurs because the normalization of the quark and gluon
fields is arbitrary.  NRQCD is the effective theory that arises in the limit $c
\rightarrow \infty$.  Having the NRQCD Lagrangian independent of $c$ (without
performing any rescaling of fields) is the motivating factor behind the
placement of factors of $c$ in the Lagrangian in Eq.~\ref{eq:16} and for the
factor of $1/c$ associated with the strong coupling in covariant derivative in
Eq.~\ref{eq:17}.

Although $\hbar$ has been set to unity, $c$ is explicit and so the dimensions
of all quantities are expressible in units of length $[x]$ and time $[t] ([E]
\sim 1/[t]$ and $[p] \sim 1/[x]$).  The gluon field has dimensions $[A] \sim
1/\sqrt{[x][t]}$ and the strong coupling $[g] \sim \sqrt{[x]/[t]}$.  The
fermion field has dimensions $[\psi] \sim 1/[x]^{3/2}$ and its mass has
dimensions $[m_Q] \sim [t]/[x]^2$.

For the fermion field, the transition from QCD to NRQCD follows the usual 
derivation of HQET.  It is rewritten as 
\begin{equation}
Q = e^{-im_{Q} c^2t} \left[1 + {i \Dslash_\perp \over 2m_Qc }+ \ldots \right] \psi,
\label{eq:18}
\end{equation}
where $\psi$ is a Pauli spinor written as a four-component object satisfying
the constraint $\gamma^0 \psi = \psi.$   The covariant derivative $D_\perp =
(0, {\bf D})$.  Using Eq.~\ref{eq:18} the part of the QCD Lagrangian density
involving $Q$ becomes
\begin{equation}
{\cal L}_{\psi} = \psi^\dagger \left[i\left({\partial\over\partial t} + 
   ig A_0^B T^B\right) + {\mbox{\boldmath$\nabla$}^2\over 2m_Q}\right] 
   \psi +\ldots,
\label{eq:19}
\end{equation}
where the ellipses denote terms suppressed by powers of $1/c$.  The leading
term is $c$ independent and corresponds to a heavy quark interacting with a
gluon potential $A_0^B$.  Among the terms suppressed by $1/c$ is the fermion
interaction term in the Lagrangian density
\begin{equation}
{\cal L}_{int} = {ig\over m_Q c} {\bf A}^B [\psi^\dagger T^B \mbox{\boldmath$\nabla$} \psi - (\mbox{\boldmath$\nabla$} \psi)^\dagger T^B \psi].
\label{eq:20}
\end{equation}
Note that ${\cal L}_{\psi}$ in Eq.~\ref{eq:19} and ${\cal L}_{int}$ in
Eq.~\ref{eq:20} both respect a heavy quark spin symmetry.  Unlike HQET there is
no heavy quark flavor symmetry because the leading term in Eq.~\ref{eq:19}
depends on the heavy quark mass.  There is another fermion interaction term
suppressed by $1/c$ involving the color magnetic field ${\bf B}_c =
\mbox{\boldmath$\nabla$} {\bf \times A}$ that breaks the spin symmetry.

It is convenient to work in Coulomb gauge $\mbox{\boldmath$\nabla$} \cdot {\bf
A} = 0$.  There are two types of transverse gluon modes that one wants to keep
in the effective field theory.\cite{labelle}  They are the potential modes
which are typically far off shell, $\partial^2 {\bf A}/\partial t^2 \ll c^2
\mbox{\boldmath$\nabla$}^2 {\bf A}$ and propogating modes where $\partial^2
{\bf A}/\partial t^2 \sim c^2 \mbox{\boldmath$\nabla$}^2 {\bf A}$.  Including
both modes is achieved by decomposing the transverse gluon field as
\begin{equation}
{\bf A}^B ({\bf x}, t) = \hat{\bf A}^B ({\bf x}, t) + \tilde{{\bf A}}^B ({\bf x}/c,t)/\sqrt{c},
\label{eq:21}
\end{equation}
where the hat is used to denote the nonpropogating potential transverse 
gluons and the tilde denote the propogating transverse gluons.  
The NRQCD Lagrangian is
\begin{equation}
L_{NRQCD} = L_A + L_{\tilde{A}} + L_{\hat A} + L_\psi,
\label{eq:23a}
\end{equation}
where
\begin{eqnarray}
L_A &=& \int d^3 x {1\over 2} (\partial_i A_0^B)^2,\\
L_{\hat A} &=& \int d^3 x (-) {1\over 2} (\partial_i \hat A_j^B)^2,\\
L_\psi &=& \int d^3 x {\cal L}_\psi,
\end{eqnarray}
and
\begin{eqnarray}
L_{\tilde{A}} &=& {1\over c} \int d^3 x \left({1\over c} 
  \partial_t \tilde{A}_i^B ({\bf x}/c,t)\right)^2 - 
  (\partial_i \tilde{A}_j^B ({\bf x}/c,t))^2 \nonumber\\*
&=& \int d^3 y (\partial_t \tilde{A}_i^B ({\bf y},t))^2 - 
  (\partial_i \tilde{A}_j^B ({\bf y},t))^2.
\label{eq:27a}
\end{eqnarray}
Eq.~\ref{eq:23a} is independent of $c$ and  corrections to it are suppressed by
factors of $1/\sqrt{c}$.  In the second line of Eq.~\ref{eq:27a} the variable
${\bf y}$ is equal to ${\bf x}/c$ and the partial derivative, $\partial_i$ is
with respect to $y^i$.  The leading order NRQCD Lagrangian in Eq.~\ref{eq:23a}
does not contain any nonabelian gluon self couplings.  

At leading order in the $1/c$ expansion there is no mixing between $\tilde{{\bf
A}}$ and $\hat {\bf A}$, and the two types of gluon fields seperately satisfy
the Coulomb gauge condition $\mbox{\boldmath$\nabla$} \cdot \hat {\bf A} =
\mbox{\boldmath$\nabla$}_y \cdot \tilde{{\bf A}} ({\bf y},t) = 0$.  In terms
suppressed by powers of $1/\sqrt{c}$ there is mixing between the zero momentum
mode of $\hat {\bf A}$ and $\tilde{{\bf A}}$.

The rescalled field $\tilde{{\bf A}}$ in Eq.~\ref{eq:21} was introduced in
Ref.~[9]. The field $\hat{{\bf A}}$ doesn't propogate and in Ref.~[9] it was
not included in Eq.~\ref{eq:21}.  From the perspective of Ref.~[9] the  terms
suppressed by factors of $1/c$ that arise from $\hat {\bf A}$ exchange are
corrections from matching full QCD onto NRQCD.  The advantage of including
$\hat{\bf A}$ in the decomposition in Eq.~\ref{eq:21} is that it automatically
performs the tree level matching.  Ref.~[10] has both $\hat {\bf A}$ and
$\tilde{{\bf A}}$ in the effective field theory.

The method I have described so far does not reproduce the power counting of
 Bodwin, Braaten, and Lepage.  Consider the transverse gluon coupling in Eq.~\ref{eq:20}.  At leading nontrivial order in $1/c$ the part involving the propogating gluons is
\begin{equation}
{\cal L}_{int} = {ig\over m_Q c^{3/2}} \tilde{{\bf A}}^B ({\bf 0},t) [\psi^\dagger T^B  \mbox{\boldmath$\nabla$} \psi - (\mbox{\boldmath$\nabla$} \psi)^\dagger T^B \psi].
\label{eq:25}
\end{equation}
In quarkonia these gluons typically have momenta of order $m_Q v^{2}/c$ and
 if we take $\Lambda_{QCD}$ to be of this order their coupling should be nonperturbative.  In other words for the propogating gluons one wants $\alpha = g^2/(4\pi c)$ of order unity.  This means that one should take
 $g \sim \sqrt{c}$ instead of $g \sim 1$ in the power counting.  Then nonabelian terms involving the propogating gluons are not suppressed by factors of $1/\sqrt{c}$.  For example, the three gluon coupling 
 is in a term in the Lagrangian of the form 
\begin{eqnarray}
L_{3~gluon} &\sim& {g\over c^{5/2}} \int d^3 x (\mbox{\boldmath$\nabla$} {\bf \tilde{A}} ({\bf x}/c,t)) {\bf \tilde{A}} ({\bf x}/c,t) {\bf \tilde{A}} ({\bf x}/c,t)\nonumber \\
&\sim& {g\over\sqrt{c}} \int d^3 y (\mbox{\boldmath$\nabla$}_y {\bf \tilde{A}} ({\bf y},t)) {\bf \tilde{A}} ({\bf y},t) {\bf \tilde{A}} ({\bf y},t),
\label{eq:26}
\end{eqnarray}
which is leading order if $g$ is of order $\sqrt{c}$.  With $g$ for propogating
gluons of order $\sqrt{c}$ the interaction in Eq.~\ref{eq:25} is suppressed by
a single factor of $1/c$, which agrees with the power counting of Bodwin,
Braaten, and Lepage.  It is worth exploring further the consistency of this
modification of the usual nonrelativistic $v/c$ power counting that occurs in
Quantum Electodynamics.

Finally I note that the color magnetic field $\tilde{{\bf B}}_c =
\mbox{\boldmath$\nabla$}_x \times {\bf \tilde{A}} ({\bf x}/c, t)$ vanishes at
leading order in $1/c$, so the spin symmetry violating term that involves
$\psi$, $\psi^{\dagger}$, and  $\tilde{{\bf B}}_c$  is suppressed by an
additional factor of $1/c$ compared with the term in Eq.~\ref{eq:25}.  This is
the reason for the prediction that $\psi$ and $\psi'$'s produced at large $p_{\perp}$ at theTevatron should be transversly aligned.\cite{cho} There are
corrections to this prediction suppressed by powers of~\cite{beneke1} $\alpha_s(2m_c)$ 
and by powers of~\cite{beneke2} $(2m_c)/p_{\perp}$.

\section{Excited Charmed Mesons in $B$ Semileptonic Decay}

The limit of QCD where the heavy quark mass goes to infinity with its
four-velocity, $v$, held fixed gives the heavy quark effective theory, HQET.  
The QCD heavy quark field $Q$ is related to its HQET counterpart $Q_v$ by
\begin{equation}
Q(x) = e^{-im_Q v x} \left[ 1 + {i\Dslash_\perp\over 2m_Q} + \ldots \right] Q_v,
\label{eq:30}
\end{equation}
where for a four vector, $X_\mu$, its perpendicular component, 
$X_{\perp\mu} = X_\mu - v \cdot X v_\mu$, satisfies $v \cdot X_\perp = 0$.  
The field $Q_v$ destroys a heavy quark of four-velocity $v$ and satisfies the 
constraint $\vslash Q = Q_v$.  Putting Eq.~\ref{eq:30} into the QCD Lagrange 
density gives 
\[
{\cal L} = {\cal L}_{HQET} + \delta {\cal L} + \ldots,\]
where the HQET Lagrange density~\cite{eichten1}
\begin{equation}
{\cal L}_{HQET} = \bar Q_v iv \cdot D Q_v
\label{eq:31}
\end{equation}
has the heavy quark spin-flavor symmetry~\cite{isgur1} and
\begin{equation}
\delta{\cal L} = {1\over 2m_Q} [O_{kin,v}^{(Q)} + O_{mag,v}^{(Q)}],
\label{eq:32}
\end{equation}
with $O_{kin,v}^{(Q)} = \bar Q_v (iD_\perp)^2 Q_v$ and $O_{mag,v}^{(Q)} = \bar
Q_v {g\over 2} \sigma_{\alpha\beta} G^{\alpha\beta} Q_v$.  The Lagrange density
$\delta {\cal L}$ contains the $1/m_Q$ corrections to the HQET
Lagrangian.\cite{eichten2}  The opertor $O_{kin,v}^{(Q)}$ breaks the flavor
symmetry but not the spin symmetry while $O_{mag,v}^{(Q)}$ breaks both the spin
and flavor symmetries.

In the $m_Q \rightarrow \infty$ limit the spin of the light degrees of freedom,
$s_\ell$, is a good quantum number~\cite{isgur2} and hadrons containing a
single heavy quark come in doublets with total spins
\begin{equation}
j_\pm = s_\ell \pm 1/2.
\label{eq:33}
\end{equation}
The ground state mesons with $Q{\bar q}$ flavor quantum numbers have
$s_\ell^{\pi_{\ell}} = 1/2^-$ giving for $Q = c$ the $D$ and $ D^*$ mesons and
for $Q = b$ the $B$ and $B^*$ mesons.  Excited mesons with $s_\ell^{\pi_{\ell}}
= 3/2^+$ have also been observed (in both the $Q = c$ and $Q=b$ cases).  For $Q
= c$ these are the $D_1$ and $D_2^*$ mesons with masses  of 2240 MeV and 2260
MeV respectively.  In the nonrelativistic constituent quark model these are $L
= 1$ orbital excitations of the ground state doublet.~\cite{rosner} These
excited charmed mesons are ``narrow'' with widths around 20 MeV.  Another
doublet with $s_\ell^{\pi_\ell} = 1/2^+$ should occur, but the mesons in it are
expected to be broad.  Expanding the meson masses in powers of $1/m_Q$ gives
the mass formula,
\begin{equation}
m_{H_{\pm}} = m_Q + \bar\Lambda^H - {\lambda_1^H\over 2m_Q} \pm 
  {n_\mp \lambda_2^H\over 2m_Q} + \ldots,
\label{eq:34}
\end{equation}
where the ellipsis denote terms suppressed by more factors of $1/m_Q$.  
In Eq.~\ref{eq:34}, $n_\pm = 2j_\pm + 1$ and $\bar\Lambda^H$ is the energy 
of the light degrees of freedom in the $m_Q \rightarrow \infty$ limit.  
The matrix elements $\lambda_1^H$ and $\lambda_2^H$ are defined by 
\begin{eqnarray} 
\lambda_1^H &=& {1\over 2v^0 m_{H_{\pm}}} \langle H_\pm (v)|
  O_{kin,v}^{(Q)}|H_\pm (v)\rangle, \\*
\label{eq:35}
\lambda_2^H &=& {\mp 1\over 2v^0 m_{H_{\pm}}n_\mp} \langle H_\pm (v)| 
  O_{mag,v}^{(Q)} | H_\pm (v) \rangle.
\label{eq:36}
\end{eqnarray}
Note that the average mass,
\begin{equation}
\bar m_H = {n_- m_{H_{-}} + n_+ m_{H_{+}}\over n_+ + n_-},
\label{eq:37}
\end{equation}
is independent of $\lambda_2$.  I use the notation $\bar\Lambda, \lambda_1,
\lambda_2$ for the ground state doublet and $\bar\Lambda', \lambda_1',
\lambda_2'$ for the excited $s_\ell^{\pi_{\ell}} = 3/2^+$ doublet.  The
measured $D^* - D$ mass splitting implies that $\lambda_2 \simeq 0.10~{\rm
GeV^2}$ and the measured $D_2^* - D_1$ mass difference implies that $\lambda'_2
\simeq 0.013~{\rm GeV^2}$.  The effects of $O_{mag}^{(Q)}$ are much smaller in
the $3/2^+$ doublet than in the ground state $1/2^-$ doublet.  Combining the
mass formulae yields expression for $\bar\Lambda' - \bar\Lambda$ and
$\lambda'_1 - \lambda_1$ in terms of hadron masses.
\begin{eqnarray}
\bar\Lambda' - \bar\Lambda &=& {m_b (\bar m_B' - \bar m_B) - 
  m_c (\bar m_D' - m_D)\over m_b - m_c} \simeq 0.39 {\rm GeV}, \\*
\label{eq:38}
\lambda_1' - \lambda_1 &=& {2m_c m_b [(\bar m_B' - \bar m_B) - 
(\bar m_D' - \bar m_D)]\over m_b - m_c} \simeq -0.23 {\rm GeV^2}.
\label{eq:39}
\end{eqnarray}

The form factors relevant for $B \rightarrow D_1 e\bar\nu_e$ and 
$B \rightarrow D_2^* e\bar\nu_e$ decay are defined by
\begin{eqnarray}
{\langle D_1 (v',\epsilon)|V^\mu | B (v)\rangle\over\sqrt{m_D m_B}} &=& 
f_{V_{1}} \epsilon^{*\mu} + (f_{V_{2}} v^\mu + f_{V_{3}} 
v^{\prime\mu})\epsilon^* \cdot v, \\*
{\langle D_1 (v',\epsilon)|A^\mu | B (v)\rangle\over\sqrt{m_D m_B}} &=& 
if_A \epsilon^{\mu\alpha\beta\gamma} \epsilon_\alpha^* v_\beta v_\gamma ', \\*
{\langle D_2^* (v',\epsilon)|A^\mu | B (v)\rangle\over\sqrt{m_{D^{*}_{2}} m_B}}
&=& k_{A_{1}}  \epsilon^{*\mu\alpha} v_\alpha + (k_{A_{2}} v^\mu + 
k_{A_{3}} v^{\prime\mu}) \epsilon_{\alpha\beta}^* v^\alpha v^\beta, \\*
{\langle D_2^* (v',\epsilon)|V^\mu | B (v)\rangle\over\sqrt{m_{D^{*}_{2}} m_B}}
&=& i k_V  \epsilon^{\mu\alpha\beta\gamma} \epsilon_{\alpha\sigma}^* 
v^\sigma v_\beta v_\gamma'.
\end{eqnarray}

The form factors $f_i$ and $k_i$ are functions of the dot product $w=v \cdot
v^{\prime}$ of the $B$ four-velocity $v$ and the charmed meson four-velocity
$v'$.  At zero recoil (i.e., $w = 1$) only $f_{V_{1}}$ can contribute to the
matrix elements above.  All other terms automatically vanish (e.g., because $v
\cdot \epsilon^* = 0$ when $v = v'$).  For $B \rightarrow D_1 e\bar\nu_e$ and
$B \rightarrow D_2^* e\bar\nu_e$ decay all of the phase space is near zero
recoil.  The entire phase space is  $1<w \lsim 1.3$.
  
$f_{V_{1}} (1)$ (i.e., the zero recoil value of the form factor $f_{V_{1}}$)
vanishes by heavy quark spin symmetry in the $m_Q \rightarrow \infty$ limit. 
In this limit, the vector and axial currents are charges of the heavy quark
spin flavor symmetry, and the $D_1$ and $D_2^*$ have $s_\ell = 3/2$ while the
$B$ and $B^*$ have $s_\ell = 1/2$.  The form factors $f_i, k_i$ are related to
a single Isgur--Wise function, $\tau$, in the infinite mass limit\cite{isgur3}:
$\sqrt{6} f_A = - (w + 1)\tau, \sqrt{6} f_{V_{1}} = (1 - w^2)\tau, \sqrt{6}
f_{V_{2}} = - 3\tau, \sqrt{6} f_{V_{3}} = (w - 2)\tau, k_V = - \tau, k_{A_{1}}
= - (1 + w)\tau, k_{A_{2}} = 0$ and $k_{A_{3}} = \tau$.  Note that the value of
$\tau (1)$ is not fixed by heavy quark symmetry.   In the infinite mass limit
$f_{V_{1}} (1) = 0$ independent of the value of $\tau(1)$.

At order $\Lambda_{QCD}/m_Q$ the form factor $f_{V_{1}}(1)$ is no longer zero. 
Recently it has been shown that at this order it can be written  in terms of
$\bar\Lambda' - \bar\Lambda$ (which is known in terms of measured masses) and
the Isgur--Wise function $\tau(w)$ evaluated at zero recoil. 
Explicity,\cite{leibovich}
\begin{equation}
\sqrt{6} f_{V_{1}} (1) = - {4\over m_c} (\bar\Lambda' - \bar\Lambda) \tau (1).
\label{eq:41}
\end{equation}
Note that the factor of four in the numerator of Eq.~\ref{eq:41} means that
this is quite a large correction.  Furthermore its importance is enhanced over
other $\Lambda_{QCD}/m_Q$ corrections since most of the phase space is near
zero recoil.  Recently ALEPH~\cite{buskulic} and CLEO~\cite{browder} have
measured (with some assumptions) the branching ratio, $Br (B \rightarrow D_1
e\bar\nu_e) = (6.0 \pm 1.1) \times 10^{-3}$.  With more experimental
information on the semileptonic decays $B \rightarrow D_1 e\bar\nu_e$ and $B
\rightarrow D_2^* e\bar\nu_e$ it will be possible to study in detail the
applicability of results based on the $\Lambda_{QCD}/m_Q$ expansion to these
decays.  This may also lead to a better understanding of how exclusive
semileptonic decays add up to the inclusive semileptonic decay rate and
influence our understanding of decays to the ground state, $B \rightarrow
D^{(*)} e\bar\nu_e$, through the application of $B$ decay sum
rules.~\cite{bjorken}

This work was supported by the U.S.\ Dept.\ of Energy under 
grant no.\ DE-FG03-92-ER~40701.

\end{document}